\documentclass[aps,prl,superscriptaddress,amsmath,amssymb,floatfix,twocolumn,showpacs]{revtex4-1}
\usepackage{times}
\usepackage{graphicx,graphics,color}
\usepackage{color}

\begin{document}

\title{Electronic Structure and Correlation Effects in PuCoIn$_5$ as Compared to PuCoGa$_5$}
\author{Jian-Xin Zhu}
\email[To whom correspondence should be addressed. \\ Electronic address: ]{jxzhu@lanl.gov}
\homepage{http://theory.lanl.gov}
\affiliation{Los Alamos National Laboratory,
Los Alamos, New Mexico 87545, USA}

\author{P. H. Tobash}
\affiliation{Los Alamos National Laboratory,
Los Alamos, New Mexico 87545, USA}

\author{E. D. Bauer}
\affiliation{Los Alamos National Laboratory,
Los Alamos, New Mexico 87545, USA}

\author{F. Ronning}
\affiliation{Los Alamos National Laboratory,
Los Alamos, New Mexico 87545, USA}

\author{B. L. Scott}
\affiliation{Los Alamos National Laboratory,
Los Alamos, New Mexico 87545, USA}

\author{K. Haule}
\affiliation{Rutgers University, Piscataway, New Jersey 08854, USA}

\author{G. Kotliar}
\affiliation{Rutgers University, Piscataway, New Jersey 08854, USA}

\author{R. C. Albers}
\affiliation{Los Alamos National Laboratory,
Los Alamos, New Mexico 87545, USA}

\author{J. M. Wills}
\affiliation{Los Alamos National Laboratory,
Los Alamos, New Mexico 87545, USA}

\begin{abstract}
Since their discovery nearly a decade ago, plutonium-based superconductors have attracted considerable interest, which is now heightened by the latest discovery of superconductivity in PuCoIn$_5$.
In the framework of density functional theory (DFT) within the generalized gradient approximation (GGA) together with dynamical mean-field theory (DMFT), we present a comparative study of the electronic structure of superconducting PuCoIn$_5$  with an expanded unit cell volume relative to its PuCoGa$_5$ cousin.
Overall, a similar GGA-based electronic structure, including the density of states, energy dispersion, and Fermi surface topology, was found for both compounds.
The GGA Pu 5$f$ band was narrower in PuCoIn$_5$ than in PuCoGa$_5$ due to the expanded lattice, resulting in an effective reduction of Kondo screening in the former system, as also shown by our DMFT calculations.

 \end{abstract}

\maketitle

{\it Introduction.~}  The itinerant-to-localized crossover of the $5f$ electrons that occurs near plutonium in the actinide series is one of the most challenging issues in condensed matter.  The cubic $\delta$ phase of plutonium metal lies closer to the localized side of this boundary, similar to the heavy actinides (Am and beyond) where the $5f$ electrons do not participate in bonding, while in $\alpha$-Pu the 5$f$ electrons are itinerant and contribute to the bonding, similar to the light actinides (Th-Np)~\cite{RCAlbers:2001}. This change in bonding leads to a 25\% larger volume in the $\delta$ phase and a low-symmetry, monoclinic crystal structure for $\alpha$-Pu, as well as a variety of unusual physical and mechanical properties~\cite{NGCooper:2000}. 
It is well established that these interesting  phenomena in elemental Pu arise from the strong electronic correlation in the 5$f$
electrons~\cite{SYSavrasov:2001,JHShim:2007,JXZhu:2007,CHYee:2010,MMatsumoto:2011,MEPezzoli:2011,Pourovskii:2006}.  With the discovery of superconductivity in PuCoGa$_5$ at $T_c$=18.5 K ~\cite{JLSarrao:2002} and later in PuRhGa$_5$ at $T_c$=8.7 K~\cite{FWastin:2003}, there is renewed interest in studying the strong electronic correlations that now also generate a transition temperature an order of magnitude higher than in their CeMIn$_5$ counterparts \cite{JJJoyce:2003,MEPezzoli:2011}.  Furthermore, superconductivity ($T_c=2.5$ K) has recently been discovered in PuCoIn$_5$~\cite{EDBauer:2011}, which has a unit cell volume 28\% larger than its itinerant superconducting  PuCoGa$_5$ cousin, similar to the volume difference between  $\alpha$-Pu and  $\delta$-Pu.  Investigating these isostructural materials provides a particularly convenient way to probe the itinerant-to-localized crossover without the complication of a drastic structural change, and to help elucidate the origin of superconductivity in the Pu-based materials. 
 
In this Letter, we present a comparative study of the electronic structure of PuCoIn$_5$ and PuCoGa$_5$.
Our calculations reveal they have the same number of Fermi surface sheets and similar band-center locations, although the details of their Fermi surface topology are sightly different. The expanded volume of PuCoIn$_5$ causes a narrower bare $5f$ band relative to that of PuCoGa$_5$. Moreover, the LDA+DMFT calculations
 show a reduction of Kondo screening in PuCoIn$_5$ relative to PuCoGa$_5$ caused by the band narrowing, which tips the balance between competing Kondo and RKKY interactions towards magnetism and localization of the $5f$ electrons in PuCoIn$_5$.



\begin{figure}[b!]
\centering\includegraphics[
width=1.0\linewidth,clip]{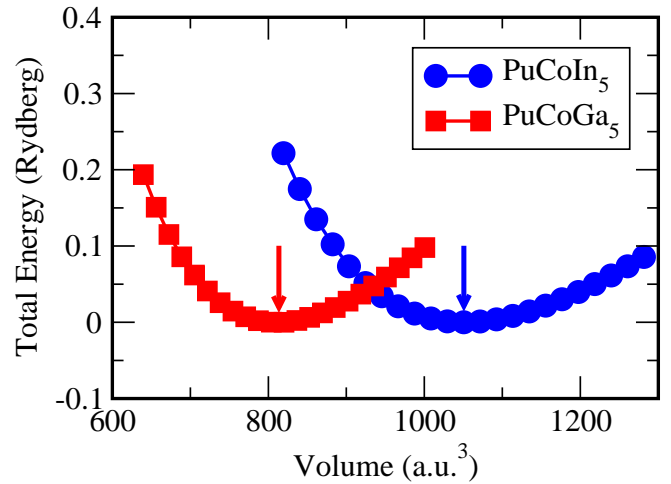}
\caption{(Color online) Calculated total energy versus volume for PuCoIn$_5$ and PuCoGa$_5$ in the paramagnetic state. The energy is shifted by 121004.4083 Rydberg for PuCoIn$_5$ and 81612.8445 Rydberg for PuCoGa$_5$, respectively. The experimentally determined volumes are shown with arrows.
}
\label{FIG:TotalEnergy}
\end{figure}

{\it Methodology.~}    We performed electronic structure calculations of PuCoIn$_5$ and PuCoGa$_5$
within the framework of density functional theory (DFT) in the
generalized gradient approximation (GGA)~\cite{JPPerdew:1996}.
Our calculations were carried out by using two relativistic band-structure methods: The full-potential linearized augmented plane wave (FP-LAPW) method as implemented in the WIEN2k code~\cite{PBlaha:2001}, and the full-potential linear muffin-tin orbital (FP-LMTO) method as implemented in the RSPt code~\cite{JMWills:2000}.
To address the 5$f$-electronic correlation issue, we used the GGA+$U$ and  GGA+DMFT~\cite{GKotliar:2006}  approximations, which are implemented in the WIEN2k code~\cite{KHaule:2010}.
For the DMFT impurity solver, we used the vertex-corrected one-crossing approximation
(OCA)~\cite{ThPruschke:1989}, which is
reasonable for the description of more localized correlated electron systems.

 \begin{figure}[t]
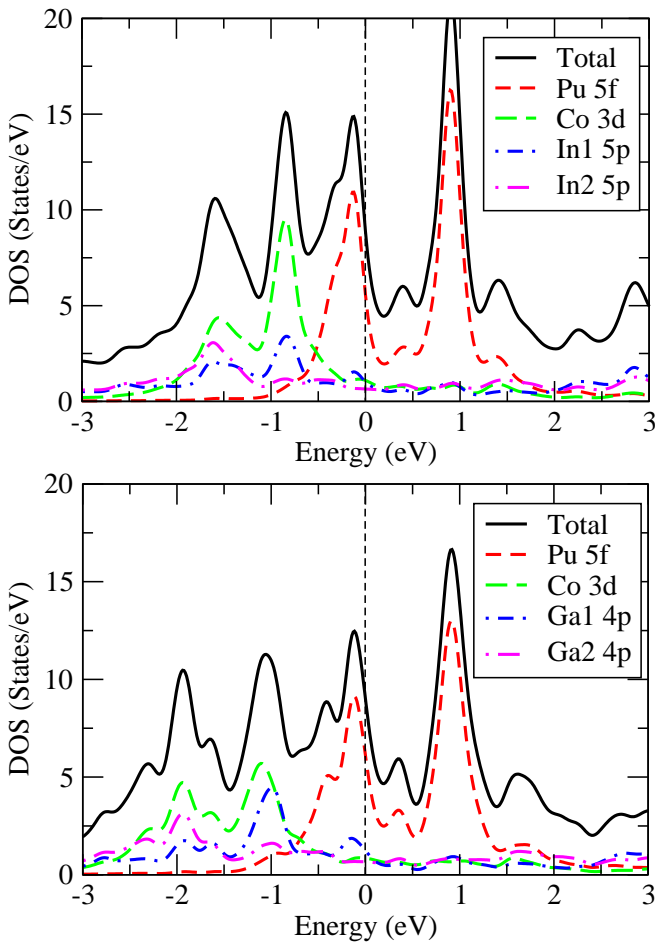

\centering
\includegraphics[width=1.0\linewidth,clip]{fig2a.eps}
\includegraphics[width=1.0\linewidth,clip]{fig2b.eps}
\caption{(Color online) Calculated GGA total and partial density of states (DOS) for PuCoIn$_5$ (top)
and PuCoGa$_5$ (bottom) in the paramagnetic state. The In 5$p$ and Ga 4$p$ DOS have been multiplied by a factor of 10 for clarity.
}
\label{FIG:LDADOS}
\end{figure}

{\it LDA bandstructure and Fermi surface topology.~}  PuCoIn$_5$ and PuCoGa$_5$ crystalize in the tetragonal HoCoGa$_5$ structure ($P4/mmm$ space group) with one internal  $z$-coordinate
for In or Ga.  We calculated the volume dependence of the GGA-based total energy with the In and Ga $z$ coordinates fixed at their experimental value of $z$(In)=0.306~\cite{EDBauer:2011} and $z$(Ga)=0.312~\cite{JLSarrao:2002}, respectively.  As shown in Fig.~\ref{FIG:TotalEnergy}, we find the theoretical equilibrium volumes to
be 1052.7 a.u.$^3$ for PuCoIn$_5$ and 811.8 a.u.$^3$ for PuCoGa$_5$, which compare reasonably well with
the experimental values of 1050.5 a.u.$^{3}$ for PuCoIn$_5$ and 820.2 a.u.$^3$ for PuCoGa$_5$.  The good agreement at the GGA level indicates that the bonding between the transition metal and ligand atoms is the dominant factor determining the equilibrium volume of these Pu-115s, with the effect of the Pu 5$f$ electron correlation secondary in this regard, which is in striking contrast to the situation for elemental Pu~\cite{SYSavrasov:2001}.
Hereafter, all calculations are performed at the experimentally determined lattice
constants~\cite{EDBauer:2011}.

Figure~\ref{FIG:LDADOS} shows the GGA total and partial density of states (DOS). Our results for the electronic structure of PuCoGa$_5$ are in good agreement with earlier 
reports~\cite{IOpahle:2003,TMehira:2003,PMOppeneer:2007}.
The two compounds exhibit somewhat similar features in the DOS. The strong spin-orbit coupling of Pu causes the 5$f$ states to be split into two manifolds or subshells, corresponding
to a total angular momentum of $j=5/2$ and $j=7/2$. The partial DOS for Pu 5$f$ orbitals shows that the
Pu 5$f_{5/2}$ states are the largest contribution at the Fermi energy, whereas
the Co 3$d$ and Ga 4$p$ or In 5$p$ orbitals have very small contributions. Furthermore,
the narrow peak corresponding to Pu 5$f_{5/2}$ is located slightly below the Fermi energy.
Both the $f_{5/2}$ and $f_{7/2}$ peaks are narrower and exhibit less structure in PuCoIn$_5$
than in PuCoGa$_5$, indicating a weakened hybridization in the former system due to the increased unit-cell volume.

\begin{figure}[b!]
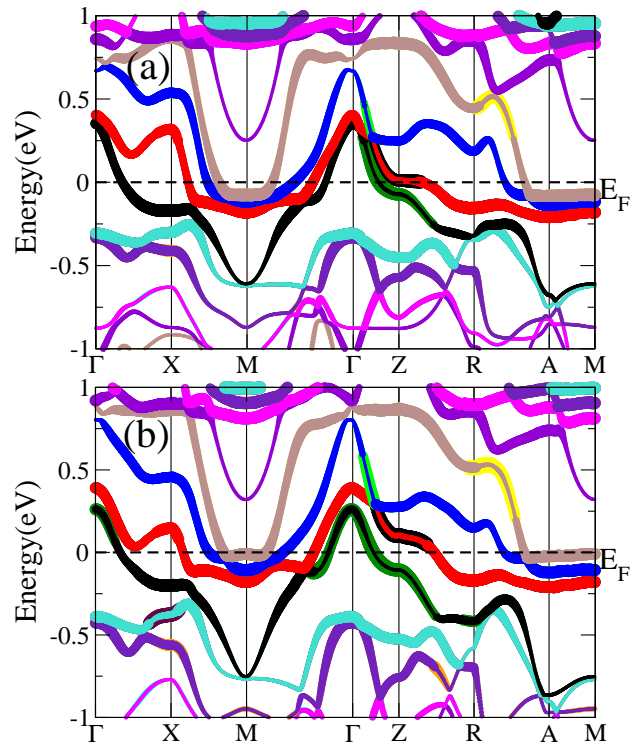

\centering
\includegraphics[
width=0.95\linewidth,clip]{fig3a.eps}
\includegraphics[
width=0.95\linewidth,clip]{fig3b.eps}
\caption{(Color online)
Energy bands of PuCoIn$_5$ (a) and PuCoGa$_5$ (b). The thickness of the lines indicates the amount of
Pu 5$f$ states present in each band.
}
\label{FIG:LDABAND}
\end{figure}

\begin{figure}[t!]
\centering
\includegraphics[width=0.45\linewidth,clip]{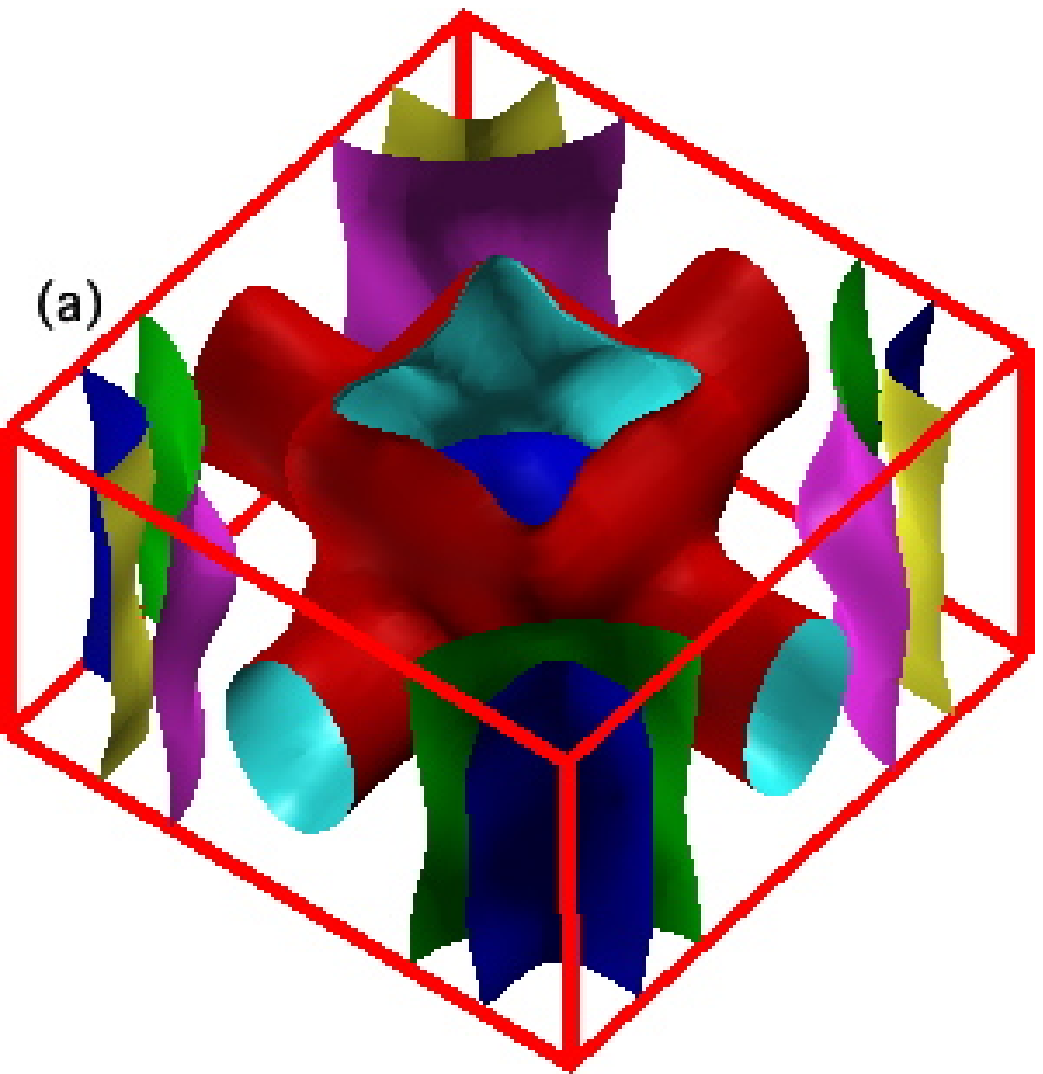}
\includegraphics[width=0.45\linewidth,clip]{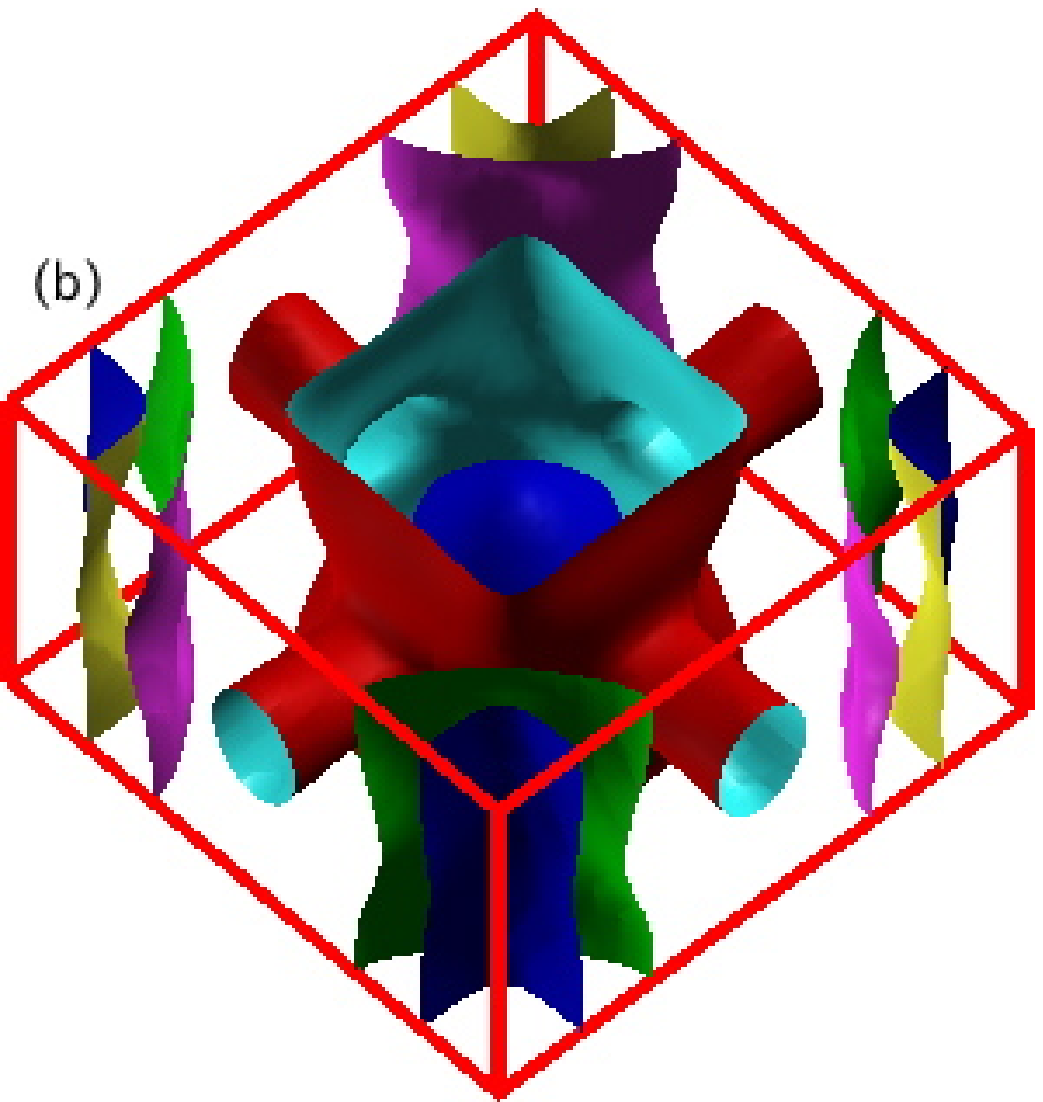}
\caption{(Color online) Calculated Fermi surface of PM PuCoIn$_5$ (a) and PuCoGa$_5$ (b).
}
\label{FIG:LDAFS}
\end{figure}

In Fig.~\ref{FIG:LDABAND}, we show the band dispersion as a function of wave vector along high-symmetry lines. The Pu 5$f$ band character is indicated by the relative thickness of each line.
The overall band structure of the two compounds is similar, and, as expected from the DOS results, the bands in the vicinity of the Fermi energy consist mainly of Pu 5$f$ states.
How these bands cut the Fermi energy determines the Fermi-surface topology.
In total, there are four bands that cut the Fermi energy, which gives rise
to four Fermi-surface sheets, as shown in Fig.~\ref{FIG:LDAFS}.
Among these four sheets, two of them are of hole character derived from the two lower bands cutting the Fermi energy and two of them of electron character derived from the two upper bands cutting the Fermi energy.
Two hole pockets are centered at the $\Gamma$ point in the zone center, while two electron pockets are centered at the $M$ point in the zone corners. At this level, the electronic structure of the Pu-115s bears some resemblance to the recently discovered Fe-based superconductors~\cite{Mazin:2008}.
Except for the small hole pocket, the Fermi surface exhibits a pronounced two-dimensional character, 
which is related to the layered crystal structure with Pu forming a square lattice in each plane.
A closer examination shows that the second hole Fermi surface (marked in red in Fig.~\ref{FIG:LDAFS}), is less square on the $z=\pm \pi$ zone face, which is due to the relative location of the four bands
with respect to the Fermi energy (see Fig.~\ref{FIG:LDABAND}).
It is worth noting that the Fermi surfaces of the three known Pu-based superconductors are all qualitatively similar.

\begin{table}
\begin{ruledtabular}
\begin{tabular}{llll}
 & \multicolumn{3}{c}{U (eV)} \\
PuCoIn$_5$  & 0.0  & 2.0   & 4.0  \\ \hline
$n_{5f}^{\text{MT}}$(SIC)            & 5.15   & 5.17   &  5.28  \\
$n_{5f}^{\text{MT}}$(AMF)          &  5.15   & 5.23   & 5.40   \\
PuCoGa$_5$  &   &   &   \\
$n_{5f}^{\text{MT}}$(SIC)                 & 5.09  & 5.05 & 5.05  \\
$n_{5f}^{\text{MT}}$(AMF)                & 5.09  & 5.11 & 5.22
\end{tabular}
\end{ruledtabular}
\caption{The Pu 5$f$ electron density within the muffin-tin sphere obtained in the GGA+$U$ approximation for both SIC and AMF methods for double counting corrections.}
\label{TABLE:n5f}
\end{table}

{\em Pu 5$f$ occupancy and correlations effects.~}
Our GGA calculations for the Sommerfeld coefficient was found to be $\gamma_{\text{GGA}}= 18\; \text{mJ}/\text{mol}\cdot \text{K}^{2}$ for PuCoIn$_5$
and $\gamma_{\text{GGA}}=21\;  \text{mJ}/\text{mol}\cdot \text{K}^{2}$ for PuCoGa$_5$. These values are smaller by a factor of about 10 and 5, respectively, than
the experimental specific coefficients, which are estimated to be 200 $\text{mJ}/\text{mol}\cdot \text{K}^{2}$~\cite{EDBauer:2011} for PuCoIn$_5$ and
80 to 116 $\text{mJ}/\text{mol}\cdot \text{K}^{2}$~\cite{JLSarrao:2002,EDBauer:2004,JDThompson:2007,PJavorsky:2005} for PuCoGa$_5$.
Although the renormalization effect is not as strong as in the Ce-115 compounds~\cite{HHegger:2000,CPetrovic:2001},
electronic correlations are still important. To understand how this affects the magnetism and superconductivity in the Pu-115s, it would
be valuable to have some insight into the Pu 5$f$ valence of these compounds. For this purpose, we have performed GGA+$U$ calculations by using two different methods for
double counting corrections: the self-interaction correction (SIC) approximation~\cite{VIAnisimov:1993} and the around-mean field (AMF) method~\cite{MTCzyzyk:1994} using an identical value of the muffin tin radius $R_{\text{MT}}=3.28\;\text{a.u.}$ for both compounds.
All calculations show that Pu 5$f$ weight remains at the Fermi level indicating some degree of mixed valent behavior. The Fermi surface was qualitatively unchanged from that presented in Fig. 4 with the addition of $U$. Table I lists the $5f$-orbitally projected electron density within the muffin tin sphere. Values close to 5 are consistent with previous estimates based on the DFT+DMFT method~\cite{MEPezzoli:2011,Pourovskii:2006}. As can be seen, only relative occupations between compounds are meaningful, since the occupation depends on the basis sets used as well as on the double-counting correction method, and is systematically larger with the AMF method than with the SIC
approximation.  However, regardless of which scheme was used, $n_{5f}^{\text{MT}}$ was found to be larger in PuCoIn$_5$ than in PuCoGa$_5$. A relatively larger value of $n_{5f}^{\text{MT}}$ indicates that the 5$f$ electron density  is more localized in PuCoIn$_5$, 
because more $5f$ electrons are pulled  inside the muffin tin.

\begin{figure}[th]
\centering\includegraphics[
width=0.9\linewidth,clip]{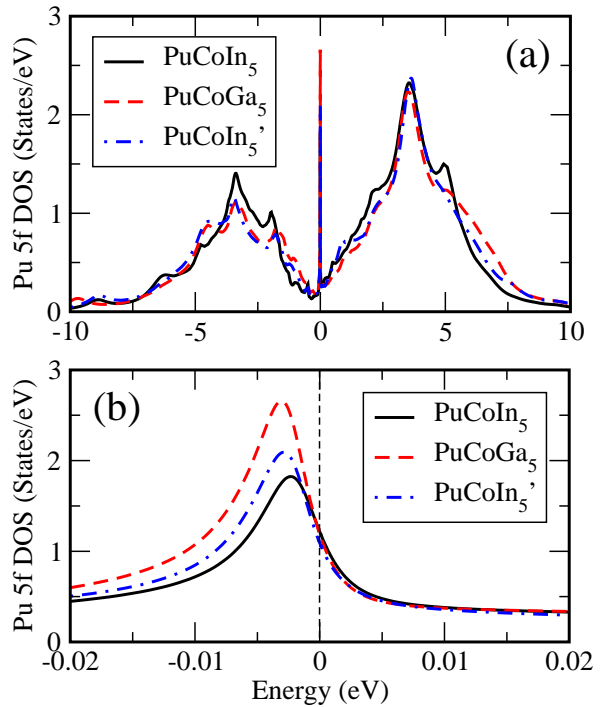}
\caption{(Color online) (a) Pu 5$f$ DOS at $T=$ 20 K in the PM PuCoIn$_5$ and PuCoGa$_5$ from the GGA+DMFT calculations.  (b) Expanded view near $E_F$. The data represented by blue line is for a hypothetical
PuCoIn$_5$ compound with a reduced unit cell volume by 20\%.
}
\label{FIG:DMFT}
\end{figure}

The observation of an enhanced specific heat coefficient and a coherence feature in transport measurements
indicate the importance of Kondo screening in these compounds, which is an effect that goes beyond what can be calculated in the framework of DFT within the local density based approximation.
In particular, the ultimate ground state of these compounds is determined
by the competition between  Kondo coupling and  magnetic exchange interactions.
To obtain a qualitative understanding of the Kondo exchange coupling in these systems, we performed GGA+DMFT calculations.
We used  $U=4\;\text{eV}$ for the Hartree component of the screened Coulomb interaction, which is consistent with previous work on elemental Pu~\cite{SYSavrasov:2001,JHShim:2007,JXZhu:2007,CHYee:2010}. The remaining Slater integrals ($F^2$, $F^4$, and $F^6$) were calculated using Cowan's atomic structure code~\cite{RDCowan:1981}
 and reduced by 30\% to account for screening.
Figure~\ref{FIG:DMFT} shows the
Pu 5$f$ partial DOS, which exhibits a three-peak structure. The two broad peaks below and above the Fermi energy correspond
to the $j=5/2$ and $j=7/2$ subshells, respectively, with an energy difference due mainly to the Hubbard $U$ and the spin-orbit coupling.  The central peak located very close to the Fermi energy
is a Kondo resonance state, which is a hallmark of quantum many-body effects. This Kondo resonance, which constitutes a strongly
renormalized quasiparticle band, is a generic feature that applies to both Pu-115 compounds. By comparing the renormalized bandwidth with the bare one, a rough estimate of the renormalization is about two orders of magnitude. This over-estimate is reasonable, since the impurity solver is based on the non-crossing type of approximation that
underestimates the Kondo screening~\cite{GKotliar:2006}. Independent of the precise value, we clearly find a narrower quasiparticle bandwidth for PuCoIn$_5$ than for PuCoGa$_5$. The fact that the quasiparticle band broadens to nearly the same amount as in PuCoGa$_5$, when the unit cell volume of PuCoIn$_5$ is reduced by 20\%, demonstrates that the reduction of Kondo screening is primarily caused by the expansion of the lattice. 

The width of the renormalized band is a characteristic energy scale that has been shown to control the maximum superconducting transition temperature in the 115 materials~\cite{NJCurro:2005}. Thus, the lower superconducting transition temperature of PuCoIn$_5$ compared with PuCoGa$_5$ may, in part, be a consequence of the reduction in Kondo screening. Future work will help elucidate the role of spin, orbital, and/or valence fluctuations for the observation of superconductivity in Pu-based compounds~\cite{RFlint:2008,KMiyake:1999}.

{\em Concluding remarks.~} We performed GGA band-structure calculations for the PuCoIn$_5$ and PuCoGa$_5$ superconductors. A similar electronic structure was found for
both compounds.  The expanded lattice in PuCoIn$_5$ relative to PuCoGa$_5$ results in a narrower bare Pu 5$f$ band width in PuCoIn$_5$ and a consequent reduction in Kondo screening.
When put in the context of the Doniach phase diagram~\cite{SDoniach:1977,LZhu:2011}, our calculations suggest that PuCoIn$_5$ is on the  weak hybridization side, while PuCoGa$_5$ is in the more strongly hybridized limit. We anticipate that a hypothetical PuCoTl$_5$ compound would possess a magneticly ordered state. To experimentally uncover the localization-delocalization transition of Pu 5$f$ electrons, PuCo(Ga,In)$_5$ alloys would be natural candidates. This study supports the notion that an expansion in lattice constant can indeed drive the Pu $5f$ electrons towards a localized state.

{\bf Acknowledgements.~}
We acknowledge useful discussions with M. Graf,  T. Durakiewicz, J. J. Joyce, and M. E. Pezzoli.
This work was performed at Los Alamos National Laboratory under the auspices of the U.S. Department of Energy, Office of Basic Energy Sciences, Division of Materials Sciences and Engineering,  and the LANL LDRD Program.

\end{document}